\documentclass[prb,twocolumn,superscriptaddress]{revtex4}
\usepackage{amsmath,amsfonts,amssymb,esint,color}
\usepackage{graphicx}

\begin{document}

\title{Floquet Weyl Phases in a Three Dimensional Network Model}

\author{Hailong Wang}

\affiliation{Division of Physics and Applied Physics, School of Physical and Mathematical Sciences, Nanyang Technological University, Singapore 637371, Singapore}

\author{Longwen Zhou}
\affiliation{Department of Physics, National University of Singapore, Singapore 117546, Singapore}

\author{Y.~D.~Chong} \email{yidong@ntu.edu.sg}

\affiliation{Division of Physics and Applied Physics, School of Physical and Mathematical Sciences, Nanyang Technological University, Singapore 637371, Singapore}

\affiliation{Centre for Disruptive Photonic Technologies, Nanyang Technological University, Singapore 637371, Singapore}

\date{ \today}

\begin{abstract}
We study the topological properties of 3D Floquet bandstructures,
which are defined using unitary evolution matrices rather than
Hamiltonians. Previously, 2D bandstructures of this sort have been
shown to exhibit anomalous topological behaviors, such as
topologically-nontrivial zero-Chern-number phases. We show that the
bandstructure of a 3D network model can exhibit Weyl phases, which
feature ``Fermi arc'' surface states like those found in Weyl
semimetals.  Tuning the network's coupling parameters can induce
transitions between Weyl phases and various topologically distinct
gapped phases.  We identify a connection between the topology of the
gapped phases and the topology of Weyl point trajectories in
$k$-space. The model is feasible to realize in custom electromagnetic
networks, where the Weyl point trajectories can be probed by
scattering parameter measurements.
\end{abstract}

\maketitle
\section{Introduction}

Topological Weyl semimetals are three-dimensional (3D) topological
phases \cite{XGWan2011,ZFang2011,LLu2013} marked by the existence of
quasiparticle states that behave as massless relativistic particles
\cite{Weyl1929}.  They were recently observed in TaAs \cite{Hasan2015,
  ZFang2015}, as well as in photonic crystals with parity-broken
electromagnetic bandstructures \cite{LLu2015}.  A realization using
acoustic lattices has also been proposed \cite{XiaoM2015}. The key
feature of these bandstructures is the existence of linear
band-crossing points, known as Weyl points, which carry topological
charges and are thus stable against perturbations.  The Weyl points
are also tied to the existence of topologically-protected ``Fermi
arc'' surface states \cite{SMYoung,Fermi_note}.

This paper introduces a method for realizing 3D Weyl phases in Floquet
lattices.  Such lattices, which include coherent wave networks and
periodically-driven lattices, are governed by evolution matrices
rather than Hamiltonians.  Previous studies have shown that Floquet
lattice bandstructures can host a variety of phases, including
topological insulator phases with protected surface states
\cite{Tanaka2010, Kitagawa2010, Lindner2011, Gong2012, MLevin2013, Titum2015}.
Most interestingly, there exist 2D ``anomalous'' Floquet insulator
phases that are topologically distinct from conventional insulators,
despite all bands having vanishing Chern numbers
\cite{Kitagawa2010,MLevin2013, Gong2014}; this is unique to Floquet
lattices, and cannot be understood in the framework of static
Hamiltonians.  There is also the intriguing possibility, raised by
Lindner \textit{et al.}, of turning a conventionally insulating
material into a Floquet topological insulator using a driving field
\cite{Lindner2011}.  Topologically non-trivial 2D Floquet systems have
been experimentally demonstrated using optical waveguide lattices
\cite{Rechtsman}, microwave networks
\cite{Liang,Liang2014,Yidong2014,Yidong2015,Gao}, and cold atom
lattices \cite{Jotzu}.  Some groups have also performed theoretical
studies of Floquet bandstructures in 3D
\cite{Lindner2013,RWang2014,Calvo2015,Narayan2015,Bomantara2016,ZouLiu2016}.
For instance, Wang \textit{et al.}~proposed using an electromagnetic
field to convert a topological insulator into a Weyl semimetal
\cite{RWang2014}; that Weyl phase, however, lacked the
Floquet-specific features that we will see in this paper.  No 3D
Floquet system has been experimentally realized to date.

Here, we describe a Floquet bandstructure arising in the context of an
experimentally feasible 3D network model \cite{ChalkerCo}.  The
network model approach differs from descriptions of Floquet systems in
terms of time-dependent Hamiltonians \cite{Tanaka2010, Kitagawa2010,
  Lindner2011, Gong2012,MLevin2013}, but cover a similar range of
phenomena (indeed, many network models can be formally mapped to
discrete-time quantum walks).  A network model is described by a
unitary matrix representing the scattering of a Bloch wave off one
unit cell \cite{Yidong2014}; the phases of the matrix eigenvalues
determine the network's ``quasienergy'' band spectrum.  Network models
originated as a tool for studying localization transitions in
disordered 2D quantum Hall systems \cite{ChalkerCo}, but have also
proven useful for describing lattices of coupled electromagnetic
waveguides, such as ring resonator lattices
\cite{Hafezi,Hafezi2,Yidong2009} and microwave networks
\cite{Yidong2014,Yidong2015,Gao}.  Notably, microwave systems have
been used to realize 2D anomalous Floquet insulators experimentally
\cite{Yidong2015,Gao}.  The 3D network models discussed in this paper
can be implemented using similar experimental setups.

The Floquet bandstructure of our 3D network exhibits Weyl phases,
which possess the usual topological bulk-edge correspondence giving
rise to Fermi arc surface states \cite{Fermi_note}.  There exist two
pairs of Weyl points: one pair in each quasienergy gap, which is the
minimum topologically allowed and achievable only in time-reversal
($\mathcal{T}$) symmetry broken systems \cite{LLu2013}. The
quasienergy spectrum is \textit{completely} gapless in the Weyl phase
(i.e., there are Bloch states at every possible quasienergy), similar
to critical Floquet bandstructures in 2D \cite{Yidong2014,Tauber}. The
Weyl points can be measured in the form of phase singularities in the
reflection coefficient, which provides an experimental route for
demonstrating their topological robustness.

Our model also provides new insights into how Weyl phases serve as
``intermediate phases'' separating topologically-distinct insulators
\cite{Murakami}.  Apart from Weyl phases, the network model also
possesses conventional insulator phases and different
$\mathcal{T}$-broken 3D ``weak topological insulator'' phases.  Each
of the weak topological insulator phases can be interpreted as a stack
of weakly-coupled 2D anomalous Floquet insulators, similar to a 3D
network model previously studied by Chalker and Dohmen
\cite{ChalkerDohmen}.  We show that the topological invariants of the
various insulator phases are related to the $k$-space windings of Weyl
point trajectories in the intermediate Weyl phases separating those
insulators; in other words, the topology of Weyl point trajectories is
tied to the topology of the Floquet bandstructures.  (A similar
relationship has previously been found in 3D quantum spin Hall systems
\cite{Murakami2008}.)  This provides a highly distinctive experimental
signature which can be probed in future realizations of the network
model.

\section{3D Network Model}
\label{sec:3D Network Model}

\begin{figure}
  \centering
  \includegraphics[width=0.48\textwidth]{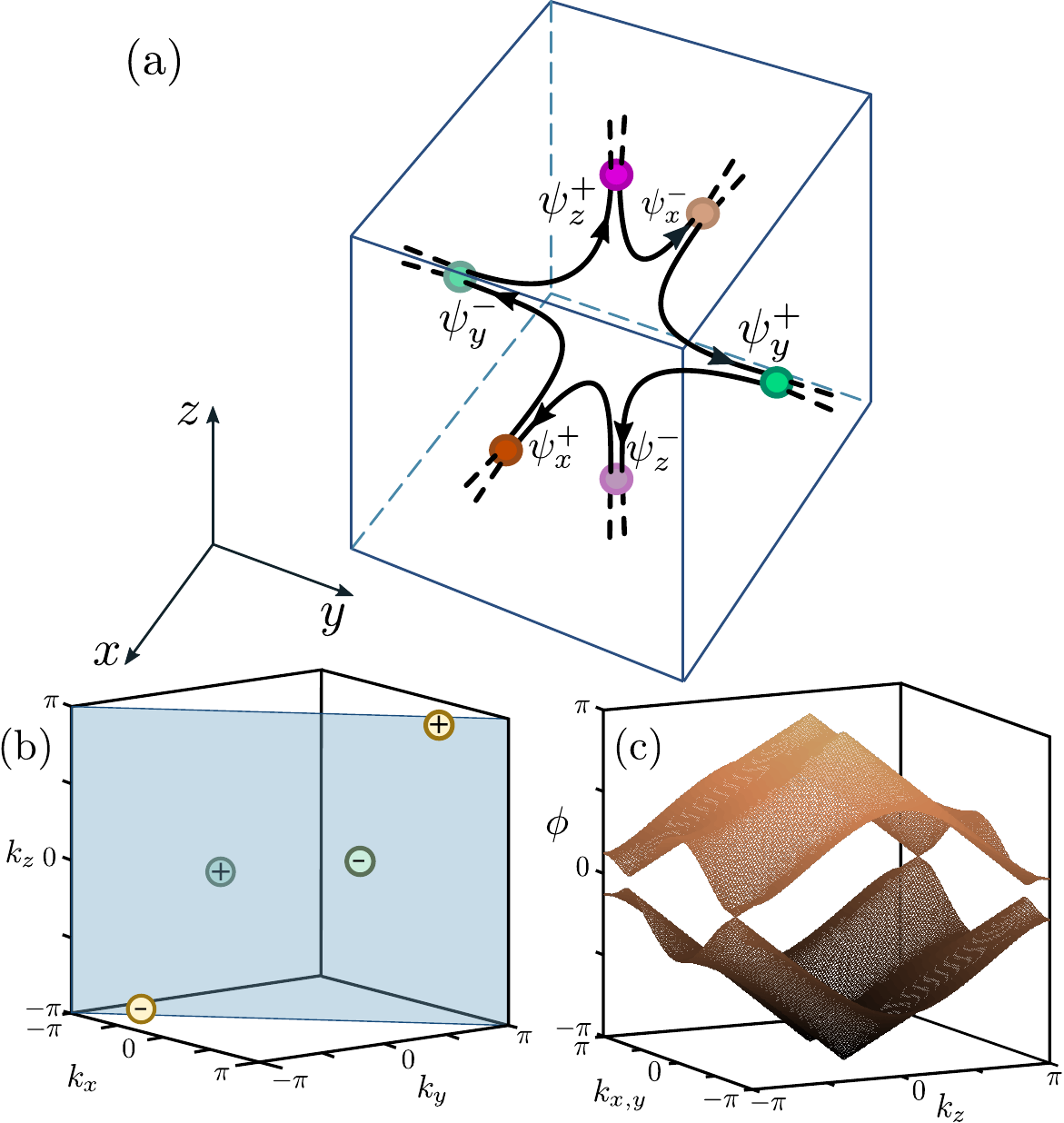}
  \caption{(Color online) (a) One unit cell of the 3D network model.
    The thick black lines are directed links, with the arrows
    indicating the directions of wave propagation.  The colored
    circles represent coupling nodes, lying at the center of each face
    of the cube.  The incident wave amplitudes are labeled
    $\psi_{\mu}^\pm$.  (b) Weyl point positions in $k$-space, for
    $\theta_x=\theta_y=\pi/8$, $\theta_z=3\pi/8$.  The Weyl points
    consist of two pairs, at $\phi = 0$ and $\phi = \pi$, all lying on
    the plane $k_x = k_y$ (shaded region).  Each Weyl point is labeled
    by its topological charge.  (c) The quasienergy bandstructure
    plotted along the $k$-space cross section $k_x = k_y$.  }
  \label{fig:network}
\end{figure}

The network model of interest is shown schematically in
Fig.~\ref{fig:network}(a).  It consists of directed links along which
waves can propagate, joined by coupling nodes.  The links and nodes
form a periodic 3D cubic lattice, with six links per unit cell.  The
links terminate at nodes located at the center of the various faces of
the cube.  Each node thus connects two incoming links, from adjacent
unit cells, to two outgoing links.  The wave amplitude at each point
along a link is given by a complex scalar.

This 3D network generalizes the 2D square-lattice network model
originally introduced by Chalker and Coddington for studying quantum
Hall systems \cite{ChalkerCo}.  For the 2D network in the
weak-coupling limit, the links in each unit cell form a closed
directed loop, analogous to the cyclotron orbits of 2D electrons in a
magnetic field.  In this 3D network, the links likewise form a
directed chiral loop in the 3D unit cell, in a way that treats the
$x$, $y$, and $z$ directions on similar footing.  In 2D, Ho and
Chalker have previously shown that the Floquet bandstructure exhibits
2D Dirac states when tuned to a critical point \cite{HoChalker};
analogously, the 3D network model exhibits Weyl phases.

Chalker and Dohmen \cite{ChalkerDohmen} have previously studied a 3D
network model consisting of stacked 2D Chalker-Coddington networks.
For weak inter-layer coupling, their network model behaves as a 3D
generalization of a quantum Hall insulator, which would nowadays be
called a $\mathcal{T}$-broken weak topological insulator.  We shall
show that our model possesses several different insulator phases,
which can be interpreted as either conventional insulators or distinct
Chalker-Dohmen insulators with different choices of weak axis.  The
relationships between these insulator phases, and the Weyl phases
separating them, will be explored in Section \ref{sec:Topological
  Invariants}.

Within a given unit cell, let $\psi_{\mu}^\pm$ denote the wave
amplitude incident on a node located along the $\mu$ direction, where
$\mu \in \{x, y, z\}$ and $\pm$ denotes the node on the positive or
negative side of axis $\mu$.  These are labeled in
Fig.~\ref{fig:network}(a).  Likewise, let $\varphi_\mu^\pm$ denotes
the wave amplitude exiting the node located on the $\pm\mu$ axis.  Let
each of the links be associated with an equal line delay $\phi/3$, so
that
\begin{equation}
  \psi_z^+ = e^{i\phi/3} \varphi_y^-, \;\;
  \psi_x^- = e^{i\phi/3} \varphi_z^+, \;\;\mathrm{etc.}
  \label{eqn:phase_delay}
\end{equation}
For an infinite network, propagating waves can be decomposed into
Bloch modes.  At each node, the incoming and outgoing wave amplitudes
are related by a $2\times 2$ unitary coupling relation
\begin{equation}
  \begin{pmatrix}
    \varphi_{\mu}^- \, e^{ik_\mu}\\ \varphi_{\mu}^+\\
  \end{pmatrix}
  =\begin{pmatrix}
  \sin\theta_\mu&i\cos\theta_\mu\\ i\cos\theta_\mu&\sin\theta_\mu\\
  \end{pmatrix}
  \begin{pmatrix}
    \psi_{\mu}^+ \\ \psi_{\mu}^- e^{ik_\mu}\\
  \end{pmatrix},
  \label{eqn:scattering}
\end{equation}
where $k_{\mu} \in [-\pi, \pi)$ is the quasimomentum, in units of the
inverse lattice period.  For convenience, we use simple unitary
$2\times 2$ coupling matrices corresponding to couplers that are
symmetric under $180^\circ$ rotations \cite{Liang,Liang2014}.  The
angle parameter $\theta_\mu \in[-\pi,\pi)$ denotes the coupling
strength along the $\mu$ direction; $\theta_\mu = 0$ corresponds to
decoupling adjacent cells.  (As discussed in Appendix \ref{3Dnetwork},
generalizing the coupling matrix to a full $2\times 2$ unitary matrix,
by including three more Euler angles, leads to trivial
translations of the bands.)

Combining Eqs.~(\ref{eqn:phase_delay})--(\ref{eqn:scattering})
yields the eigenvalue problem
\begin{equation}
  U(k) \Psi(k) = e^{-i\phi(k)} \Psi(k),
  \label{eqn:scattering2}
\end{equation}
where $\Psi = [\psi_{z}^{+}, \psi_{z}^{-}]^{\text{T}}$, and $U$ is a
$2\times2$ unitary matrix depending on $\{k_\mu,\theta_\mu\}$.
Details are given in Appendix \ref{3Dnetwork}.

We regard $\phi(k)$ as a \textit{quasienergy}
\cite{HoChalker,Liang,Yidong2014}, analogous to the band energy of a
crystal except that it is an angle with $2\pi$ periodicity.  If the
network is realized using electromagnetic waveguides for the links
\cite{Liang,Yidong2014,Yidong2015,Gao}, the quasienergy is fixed by
the phase delay of the waveguides; this is analogous to probing one
single frequency in a photonic crystal, or one energy (e.g.~the Fermi
level) in an electronic system.  Depending on the experimental
realization, it may also be possible to vary $\phi$ continuously,
e.g.~by tuning the operating frequency to alter the phase delay in the
waveguides.  Alternatively, $U(k)$ can also be derived as the Floquet
operator of a 3D discrete-time quantum walk, as described in Appendix
\ref{sec:quantumwalk}.  In the quantum walk context, $\phi(k)$
describes the temporal periodicity of a Floquet eigenmode
\cite{Tanaka2010, Kitagawa2010, Lindner2011, Gong2012, MLevin2013};
it is usually difficult to select a specific quasienergy, and instead
one excites a specific quasimomentum or lattice position, which
generates a superposition of Floquet eigenmodes \cite{Rechtsman}.

By analytically diagonalizing $U$, we find that the 3D parameter space
$(\theta_x,\theta_y,\theta_z)$ is divided into three sets of phases:
(a) an octahedron with vertices at
$(\pm\frac{\pi}{2},\pm\frac{\pi}{2},\pm\frac{\pi}{2})$, corresponding
to a conventional insulator; (b) eight tetrahedra with bases lying on
each face of the (a) octahedron and vertices at $\theta_\mu = \pm
\pi$, corresponding to gapless Weyl phases; and (c) the regions
outside (a) and (b), which form three octahedra (modulo $\pi$ in the
$\theta_\mu$ parameters), and turn out to be weak topological
insulators.  We first focus on the Weyl phases; the insulator phases
will be explored in Section \ref{sec:Topological Invariants}.

In each Weyl phase, the bandstructure is \textit{completely} gapless:
the quasienergy bands are connected by simultaneous band-crossing
points, such that there is no gap at any quasienergy $\phi \in
[-\pi,\pi)$.  (A similar phenomenon has previously been seen in
  critical 2D Floquet bandstructures \cite{Yidong2014,Tauber}.)  There
  are two pairs of band-crossing points. One pair occurs at
  quasienergy $\phi=\pi$ and $k = \pm \bar{k}$; the analytic
  expressions for $\{\bar{k}_x,\bar{k}_y,\bar{k}_z\}$ in terms of
  $\{\theta_x,\theta_y,\theta_z\}$ are given in Appendix
  \ref{3Dnetwork}.  The other pair occurs at quasienergy $\phi=0$ and
  $k = \pm(\bar{k} - \pi)$.  Fig.~\ref{fig:network}(b) shows the
  $k$-space positions of the band-crossing points (Weyl points) for
  $\theta_x=\theta_y=\pi/8$, $\theta_z=3\pi/8$, and
  Fig.~\ref{fig:network}(c) shows the quasienergy bandstructure.

Near the Weyl point at $\bar{k}$, we can expand the evolution matrix
as $U(\bar{k} + \kappa) = e^{-iH_{\text{eff}}}$, where
\begin{equation}
  H_{\text{eff}}(\kappa)=\nu_{ij} \kappa_{i} \sigma_j,\;\;\; \nu_{ij}\in\mathbb{R}.
\end{equation}
This is the Weyl Hamiltonian describing a massless relativistic
particle; similar expansions can be performed around each of the other
three Weyl points.  The $\nu$ matrix satisfies
\begin{multline}
  \det[\nu_{ij}]=-\sin\theta_z\sin{\bar
    k_z} \\ \times \left(\cos\theta_y\sin\theta_z\cos{\bar
    k_z} +
  \cos\theta_z\sin\theta_y\cos{\bar k_y}\right) \\ \times
  \left(\cos\theta_x\sin\theta_z\cos{\bar
    k_z}+\cos\theta_z\sin\theta_x\cos{\bar k_x}\right).
  \label{eqn:detnu}
\end{multline}
Each Weyl point is associated with a topological charge, consisting of
the Berry flux of a band integrated over a $k$-space sphere
surrounding the point.  This takes values
${\text{sgn}}(\det[\nu_{ij}])=\pm1$, representing the chirality of the
Weyl particles \cite{Hosur2013}.  (It is also necessary to choose
which band's Berry flux to put into the calculation; we adopt the
convention of using the band \textit{below} each Weyl point.)  Each
two Weyl points at the same quasienergy have opposite chiralities,
consistent with the principle that the total chirality sums to zero
(the Nielsen-Ninomiya theorem) \cite{Nielsen1981a,Nielsen1981b}.

The locations and chiralities of the Weyl points are constrained by
the symmetries of $U(k)$.  According to Kitagawa \textit{et al.}'s
classification of Floquet operator symmetries \cite{Kitagawa2010}, a
system is time-reversal ($\mathcal{T}$) symmetric if $U^\dagger=
\mathcal{Q} \, U^{*} \, \mathcal{Q}^\dagger$, where $\mathcal{Q}$ is
some unitary operator; and it is particle-hole symmetric if $U =
\mathcal{P} U^{*} \mathcal{P}^\dagger$ for some unitary $\mathcal{P}$.
Our network model breaks $\mathcal{T}$ due to the directed nature of
the network (Fig.~\ref{fig:network}), and this can be verified from
the fact that $U^\dagger$ and $U^{*}$ have different spectra.  On the
other hand,
\begin{equation}
  U(k)=\sigma_y U^*(k) \sigma_y,
  \label{particle_hole_symmetry}
\end{equation}
corresponding to a particle-hole symmetry with $\mathcal{P}=
i\sigma_y$.  For every eigenstate at quasienergy $\phi$ and momentum
$k$, there is an eigenstate at $-\phi$ at the same $k$.  Thus, Weyl
points can only occur at $\phi=0$ and $\phi=\pi$.  We can also
conclude that chiral symmetry is broken \cite{Ryu2010}.

The network model also satisfies inversion symmetry:
\begin{equation}
  U(-k)=\sigma_x U(k)\sigma_x.
\end{equation}
This explains why the Weyl points occur in pairs, at $k$ and $-k$,
with opposite chiralities.

There is one more interesting symmetry of $U(k)$:
\begin{equation}
  U(k) = -\sigma_z\, U\!\left(k + [\pi,\pi,\pi]\right)\, \sigma_z.
  \label{k_symmetry}
\end{equation}
Thus, for every eigenstate at quasienergy $\phi$ and momentum $k$,
there is an eigenstate at $\phi+\pi$ and $k + [\pi,\pi,\pi]$.  This,
together with the particle-hole symmetry, guarantees that Weyl points
at $\phi=0$ and $\phi=\pi$ occur simultaneously.  (A similar
phenomenon occurs in the 2D network model \cite{Tauber}, though in 2D
the gap-closings occur only at critical points.)  Thus, the Weyl point
at $\phi=\pi$ and $k=\bar{k}$ has the same chirality as the one at
$\phi = 0$ and $k = -(\bar{k} - [\pi,\pi,\pi])$, and likewise for the
other pair.

Eqs.~(\ref{particle_hole_symmetry})--(\ref{k_symmetry}) hold for a
network in which the node couplings have the highly symmetric form
given by Eq.~(\ref{eqn:scattering}).  Adopting more general node
couplings will shift the bandstructure in $\phi$ and/or $k$, as
discussed in Appendix \ref{3Dnetwork}.  This correspondingly modifies
the symmetry relations; in particular, the Weyl points may move away
from the special quasienergies ($\phi = 0$ and $\phi = \pi$) and
$k$-space plane on which they were previously constrained, without
lifting the band degeneracies of the Weyl points themselves.

\section{Fermi arc surface sates}
\label{sec:Fermi_arc}

Weyl points in a bulk bandstructure have a topological correspondence
with the existence, in a truncated lattice, of ``Fermi arc'' surface
states.  At a given energy, these surface states lie along an open arc
in the projected 2D $k$-space, connecting a pair of Weyl cones
\cite{ZFang2011,SMYoung}.  This is fundamentally different from
ordinary 2D bands, which must form closed $k$-space loops, and is
possible only because the surface states lie on the boundary of a 3D
bulk.

\begin{figure}
\centering
\includegraphics[width=1.0\linewidth]{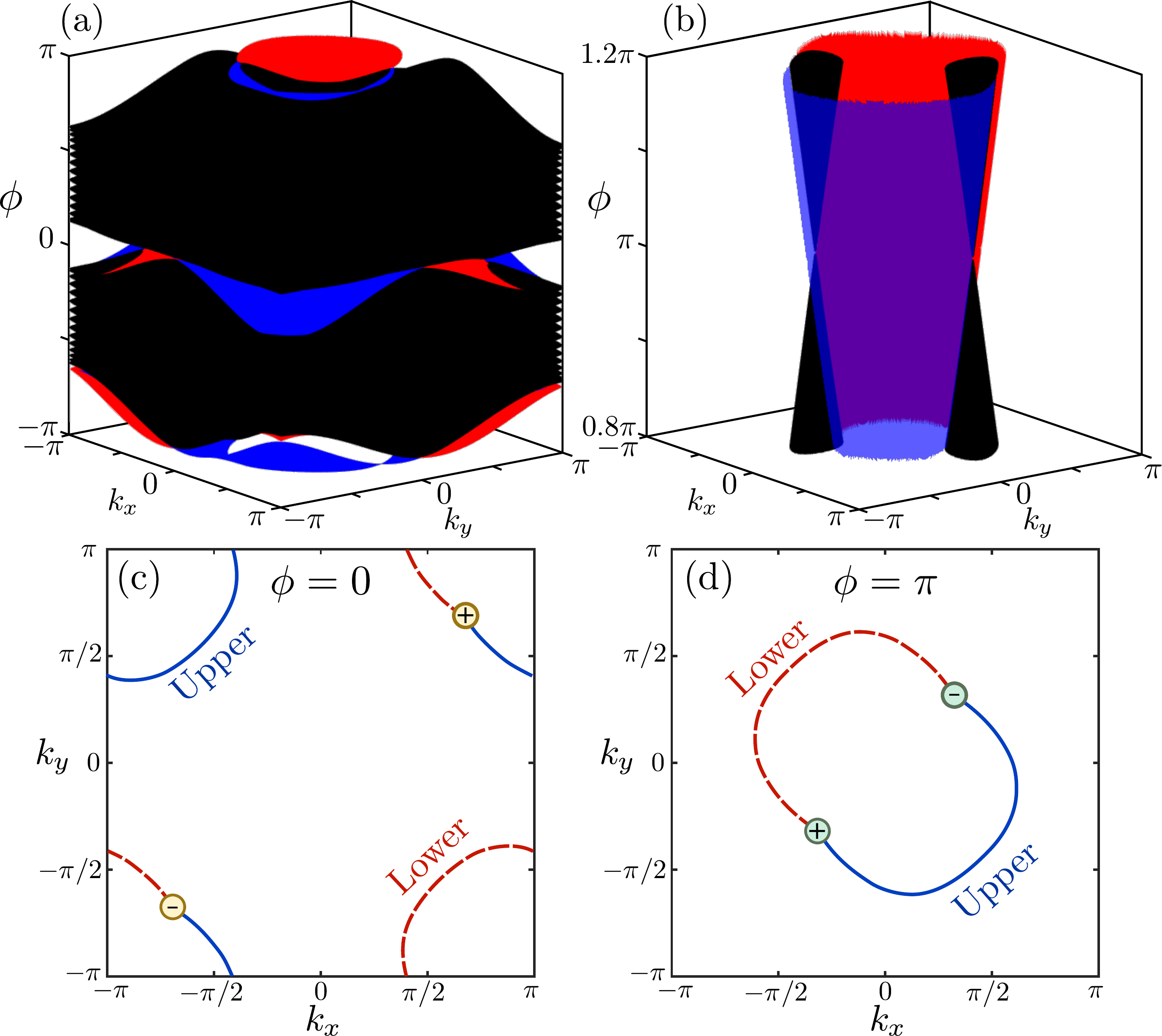}
\caption{(Color online) (a) Quasienergy bandstructure for the 3D
  network in a slab geometry with thickness $N=16$. The coupling
  parameters are $\theta_x=\theta_y=\pi/8$, $\theta_z=3\pi/8$. (b)
  Close-up view of the bandstructure near $\phi=\pi$, showing bulk
  Weyl cones and Fermi arc surface states.  (c)--(d) Cross sections of
  the bandstructure at (c) $\phi=0$, and (d) $\phi=\pi$; circles
  indicate Weyl points with positive ($+$) and negative ($-$) topological
  charges, calculated using the Berry flux on the band below each Weyl
  point. }
\label{fig:3D_surface_spectrum}
\end{figure}

To confirm that the gapless phases of the network model are Weyl
phases, we have numerically calculated the surface states and observed
the presence of Fermi arcs.  Details of the calculation are given in
Appendix \ref{Tmatrix}. Essentially, we impose a ``slab'' geometry by
truncating the 3D network to $N$ periods along $z$, while keeping it
infinite and periodic in $x$ and $y$ directions with quasimomentum
$(k_x, k_y)$.  Wave amplitudes in adjacent cells are related by
\begin{equation}
  M
  \begin{pmatrix}
    \varphi_{z}^{-}(j)\\ \psi_{z}^{-}(j)\\
  \end{pmatrix} =
  \begin{pmatrix}
    \varphi_{z}^{-}(j+1)\\ \psi_{z}^{-}(j+1)\\
  \end{pmatrix},
  \label{eqn:transfer2}
\end{equation}
where $j$ is the position index along $z$, and $M$ is a transfer
matrix that depends on $(k_x,k_y)$, $\phi$, and $\{\theta_\mu\}$.  We
impose Dirichlet boundary conditions by setting $\psi_z^-=\varphi_z^-$
on cell $j=1$ (the bottom surface), and $\psi_z^+=\varphi_z^+$ on cell
$j=N$ (the upper surface).  We then calculate the bandstructure by
searching for values of $\phi$ consistent with the boundary
conditions, for each $(k_x, k_y)$ in the 2D Brillouin zone.

The results are shown in Fig.~\ref{fig:3D_surface_spectrum}.
Fig.~\ref{fig:3D_surface_spectrum}(a) shows the entire bandstructure,
with bulk bands plotted in black and the surface state bands for the
upper (lower) surface plotted in blue (red).
Fig.~\ref{fig:3D_surface_spectrum}(b) represents the portion of the
bandstructure near $\phi = \pi$, clearly showing the existence of two
Weyl cones, to which the surface states are attached along Fermi arcs.
Fig.~\ref{fig:3D_surface_spectrum}(c) and (d) plot cross-sections
taken at the Weyl point quasienergies $\phi = 0$ and $\phi = \pi$,
showing how each Fermi arc connects a pair of Weyl points with
opposite chiralities.

In the condensed matter setting, Fermi arc states have been predicted
to exhibit novel behaviors, including quantum oscillations in
magnetotransport and quantum interference effects in tunneling
spectroscopy \cite{Hosur2013,Burkov2015}.  If the present model is
implemented using a electromagnetic network, however, most of the
previously-identified experimental signatures will be unavailable due
to the absence of a Fermi level.  In the next section, we discuss a
different approach to probing the topological structure of the Weyl
phase, based on reflection measurements.

\section{Topological Invariants}
\label{sec:Topological Invariants}

In order to clarify the nature of the gapped phases, we seek to
formulate topological invariants for the various phases of the network
model.  A standard way to characterize gapped 3D phases is to
calculate a triplet of Chern numbers obtained by integrating the Berry
flux across sections of the 3D Brillouin zone corresponding to the
$yz$, $zx$ and $xy$ planes
\cite{Avron1983,Moore2007,HasanRMP2010}. For example, integrating
across the $yz$ plane gives
\begin{equation}
  \nu_{x}^n = \frac{1}{2\pi}\oiint B_n(k)\cdot ds_{yz},
  \label{eqn:Chern}
\end{equation}
where $B_n(k) = \nabla_{k} \times A(k)$ is the Berry curvature and
$A_n(k) = i\langle \Psi_{nk}|\nabla_{k}|\Psi_{nk}\rangle$ is the Berry
connection evaluated on band $n$, computed using bulk Bloch
functions. By the Gauss-Bonnet theorem, any such integral taken over a
closed 2D surface must give an integer.

For all the gapped phases of the network model, we find that $\nu_x^n
= \nu_y^n = \nu_z^n = 0$, which ordinarily indicates that these phases
are topologically trivial.  However, Chern number invariants can be
misleading in characterizing Floquet bandstructures.  In 2D, it has
previously been shown that an anomalous Floquet insulator phase can
exist which is topologically non-trivial but has zero Chern numbers in
all bands \cite{Kitagawa2010,MLevin2013,Yidong2014}, with the
topological non-triviality verified by the existence of protected edge
states.  Roughly speaking, the anomalous Floquet insulator phase
arises by applying band inversions in every gap of the conventional
insulator, including the gap which ``wraps around'' the quasienergy
$\phi$ (which is an angle variable).  This phenomenon is unique to
Floquet systems, and cannot occur with static Hamiltonians.  As we now
argue, a similar phenomenon occurs in the 3D network model, so that
there are actually four topologically distinct gapped phases which
cannot be distinguished using Chern number triplets.

To characterize these gapped phases, we introduce a ``topological
pumping'' process \cite{Brouwer1998,Brouwer2011,Fulga,Fulga2015}.
As originally introduced by Brouwer \textit{et al.}
\cite{Brouwer1998}~for 2D systems, the idea is to roll a lattice into
a cylinder with tunable phase slip $k$ along the azimuthal direction,
and attach scattering leads to the ends of the cylinder.  When the
cylinder length greatly exceeds the attenuation length for the bulk
gap, the scattering matrix reduces to two unitary blocks, $r_\pm$,
which are the reflection matrices off each end.  The topological
invariant is the integer winding number of the reflection phase,
\begin{equation*}
  -i\ln\left[\det{r_\pm}\right],
\end{equation*}
as $k$ advances through $2\pi$.  This concept can be applied directly
to 2D network models, and has been used to experimentally distinguish
between topological phases of a microwave network \cite{Yidong2015}.

We apply the topological pumping procedure to the 3D network model in
the following way: consider the slab geometry from Section
\ref{sec:Fermi_arc}, with $N$ cells stacked in the $z$ direction and
wave-vector $(k_x,k_y)$.  Using the transfer matrix relation in
Eq.~(\ref{eqn:transfer2}), we derive a scattering relation
\begin{equation}
  \mathcal{S}(k_x, k_y, \phi, N)
  \begin{pmatrix}
    \psi_{z}^{+}(N)\\\psi_{z}^{-}(1)\\
  \end{pmatrix} =
  \begin{pmatrix}
    \varphi_{z}^{+}(N)\\ \varphi_{z}^{-}(1)\\
  \end{pmatrix},
  \label{eqn:scattering3}
\end{equation}
which relates the wave amplitudes on the upper and lower surfaces.  In
the large-$N$ limit, if $\phi$ falls into a quasienergy gap, the
scattering matrix becomes purely reflecting:
\begin{equation}
  \mathcal{S} \rightarrow
  \begin{pmatrix}
    e^{i\omega_+}&0\\0&e^{i\omega_-}
  \end{pmatrix}.
  \label{eqn:Sslab}
\end{equation}
We then pick one of the diagonal entries, say the one corresponding to
reflection off the lower surface, and determine how its phase winds
with $k_x$ and $k_y$.  (The other surface will have the opposite
windings.)  This process can be repeated for slabs taken perpendicular
to the $x$ and $y$ directions.

\begin{figure}
\centering
\includegraphics[width=0.48\textwidth]{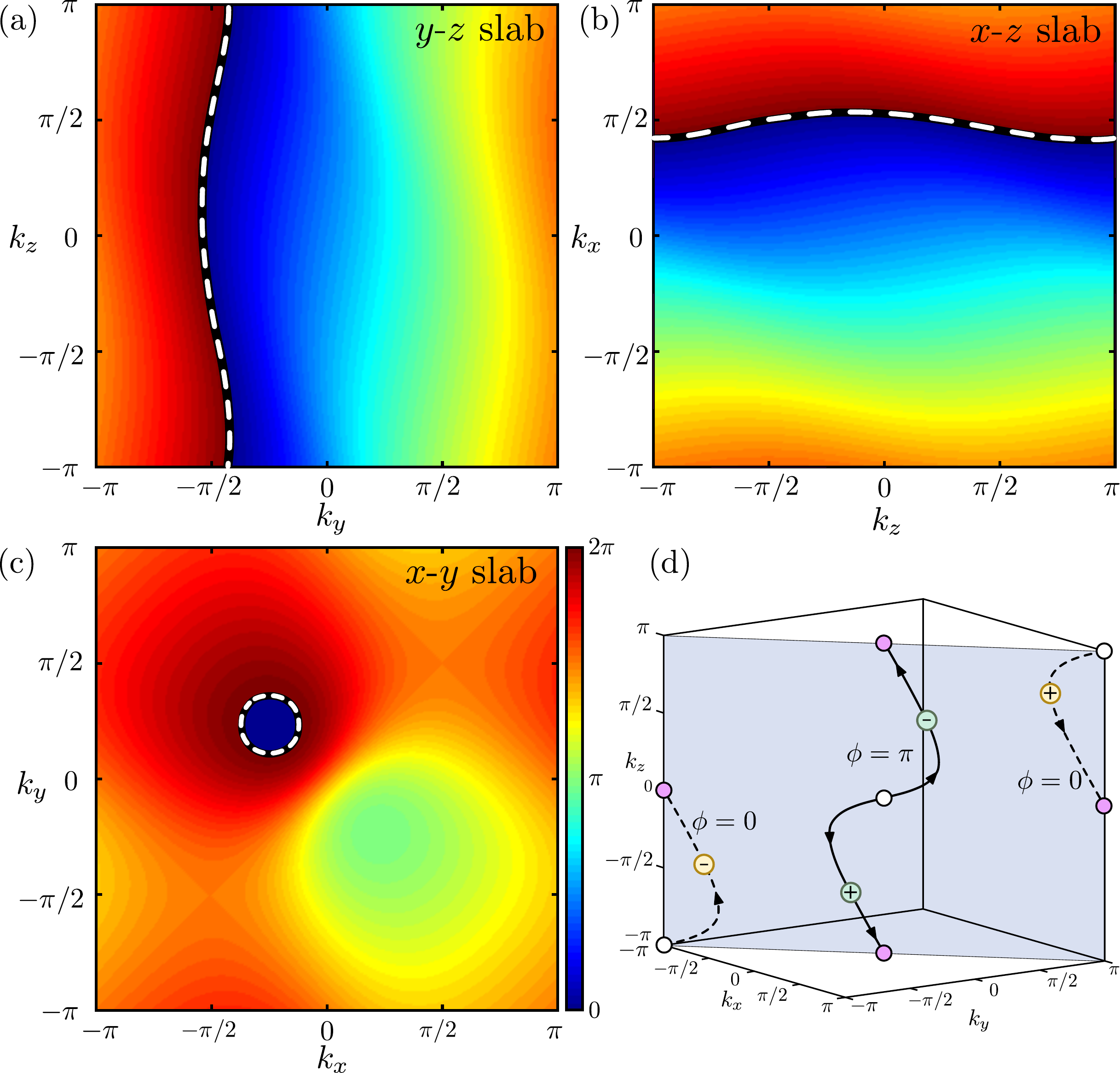}
\caption{(Color online) (a)--(c) Heat maps of the reflection phase
  $\omega_-$ in a gapped phase.  We take $\theta_x=\theta_y=3\pi/8$
  and $\theta_z=\pi/8$ [green star in
    Fig.~\ref{fig:3D_phase_diagram}(a)], and quasienergy $\phi = \pi$.
  The slab is taken in the (a) $y$-$z$, (b) $x$-$z$, and (c) $x$-$y$
  planes.  Dashed lines are contours of $\omega_- = 0$. (d) Trajectory
  of the Weyl points when traversing a Weyl phase from a $(0,0,0)$
  conventional insulator phase to a $(0,0,1)$ weak topological
  insulator phase, along the path indicated in
  Fig.~\ref{fig:3D_phase_diagram}(a). }
\label{fig:3D_reflection_coeff1}
\end{figure}

Fig.~\ref{fig:3D_reflection_coeff1}(a)--(c) shows the reflection phase
$\omega_-$ versus $k_\mu$, using the parameters
$\theta_x=\theta_y=3\pi/8$ and $\theta_z=\pi/8$.  The $y$-$z$ slab
exhibits winding number $+1$ in the $k_y$ direction, and the $x$-$z$
slab exhibits winding number $-1$ in the $k_x$ direction.  The winding
in the $k_z$ direction is zero for these slabs, and for the $x$-$y$
slab it is zero in both $k_x$ and $k_y$.  The reflection phase on the
upper surface, $\omega_+$, has opposite windings from what is shown
here.

\begin{figure}
  \centering
  \includegraphics[width=0.48\textwidth]{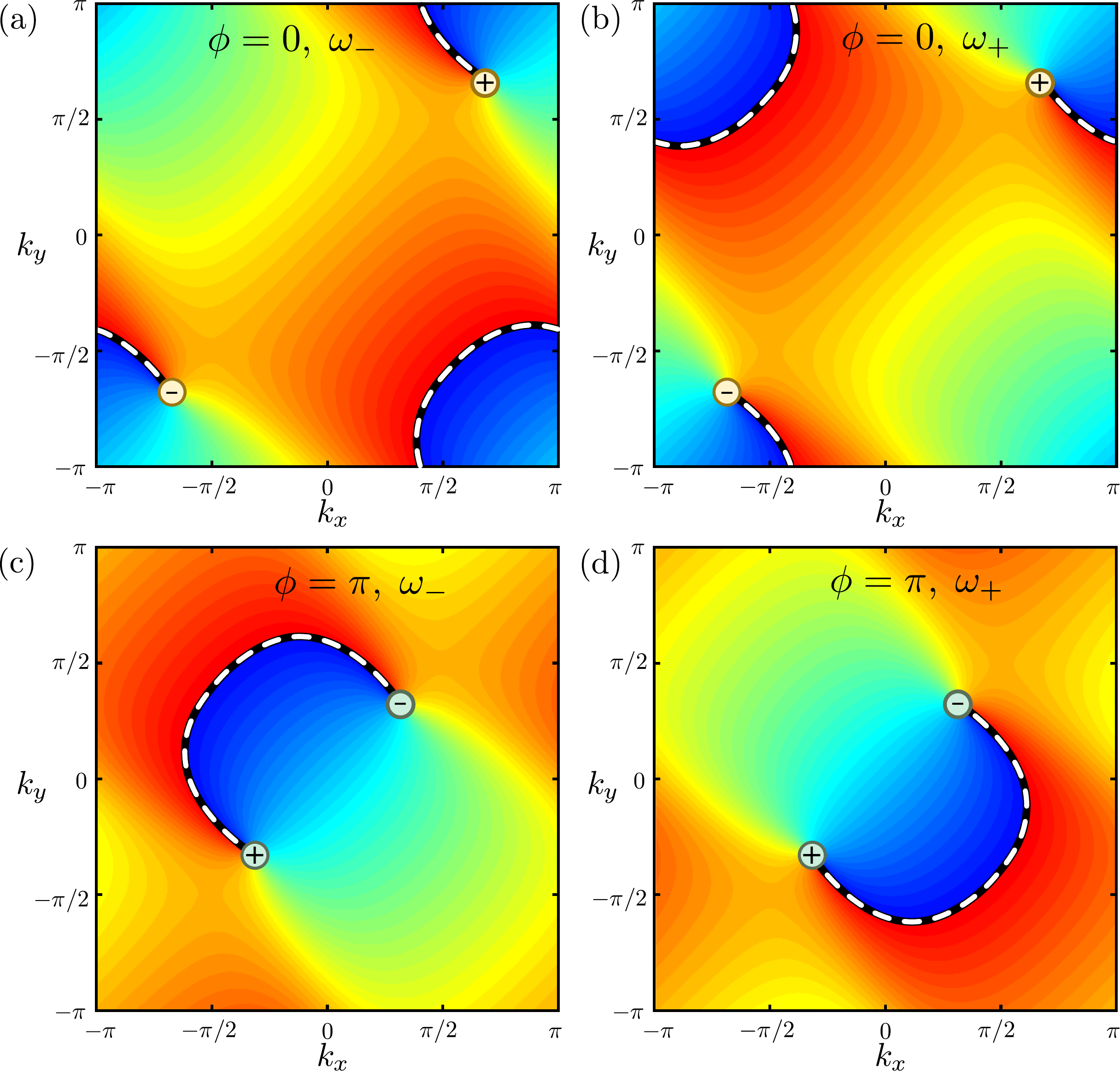}
  \caption{(Color online) Heat maps of the reflection phases
    $\omega_\pm$ in a Weyl phase.  We take $\theta_x=\theta_y= \pi/8$
    and $\theta_z=3\pi/8$ [blue star in
      Fig.~\ref{fig:3D_phase_diagram}(a)], and an $x$-$y$ slab.  In
    (a) and (b), we fix the quasienergy at $\phi = 0$ and plot (a) the
    reflection phase $\omega_-$ off the lower surface, and (b) the
    reflection phase $\omega_+$ off the upper surface.  In (c) and
    (d), we fix the quasienergy at $\phi = \pi$ and plot (c)
    $\omega_-$ and (d) $\omega_+$.  The projection of each Weyl point
    onto the plane coincides with a phase singularity; the circles
    labeled $\pm$ indicate the Weyl point's topological charge.  The
    dashed lines are contours of $\omega_\pm = 0$.}
\label{fig:3D_reflection_coeff}
\end{figure}

The topological pumping process can also be carried out in the Weyl
phase.  We can fix $\phi$ at quasienergy $0$ or $\pi$, corresponding
to one pair of Weyl points, and plot the reflection phase map.
Fig.~\ref{fig:3D_reflection_coeff} shows the results for
$\theta_x=\theta_y= \pi/8$ and $\theta_z=3\pi/8$, and an $x$-$y$ slab.
At two specific $k$ points, corresponding to the projection of the
Weyl points onto the plane, the system is gapless, and
Eq.~(\ref{eqn:Sslab}) breaks down since the slab is not purely
reflecting.  Everywhere else in the 2D Brillouin zone, the system is
gapped and Eq.~(\ref{eqn:Sslab}) holds, so that $\omega_\pm$ is
well-defined.  In the resulting reflection phase map, the $k$ points
corresponding to the Weyl point projections are associated with phase
singularities; the phase winds by $\pm 2\pi$ around each singularity,
and is ill-defined at the singularity itself.  For $\omega_+$
(reflection off the upper surface), the winding number of each phase
singularity---using right-handed coordinates, and treating
anti-clockwise winding as positive---is equal to the Weyl point's
topological charge as computed from its Berry flux (Section
\ref{sec:3D Network Model}).  For $\omega_-$ (the lower surface), the
winding numbers and topological charges have opposite signs.  This
holds at both Weyl point quasienergies.

The reflection phase singularities are intimately related to the
phenomenon of Fermi arcs.  The phase singularities can be joined by
equal-phase contours, and the contours for $\omega_\pm = 0$ (shown as
dashes in Fig.~\ref{fig:3D_reflection_coeff}) exactly match the Fermi
arc surface states plotted in
Fig.~\ref{fig:3D_surface_spectrum}(c)--(d).  This is because the Fermi
arc surface states arise from Dirichlet boundary conditions that are
equivalent to setting $\omega_\pm = 0$ in the topological pump.

Based on the above discussion, the windings of the reflection phase in
each gapped phase can be characterized by an integer triplet, the
``pumping invariant'' $\vec{n} = (n_x, n_y, n_z)$.  This is defined as
follows: for the slab with normal vector $\hat{s}$,
\begin{equation}
  (\text{winding in direction}\;
  \hat{\mu}) \; =\; \pm \, (\hat{s} \times \vec{n})\cdot \hat{\mu},
\end{equation}
where $\pm$ corresponds to the choice of $\omega_\pm$ (i.e., upper or
lower surface).  Both quasienergy gaps give the same value for $(n_x,
n_y, n_z)$.

\begin{figure}
  \centering
  \includegraphics[width=1.0\linewidth]{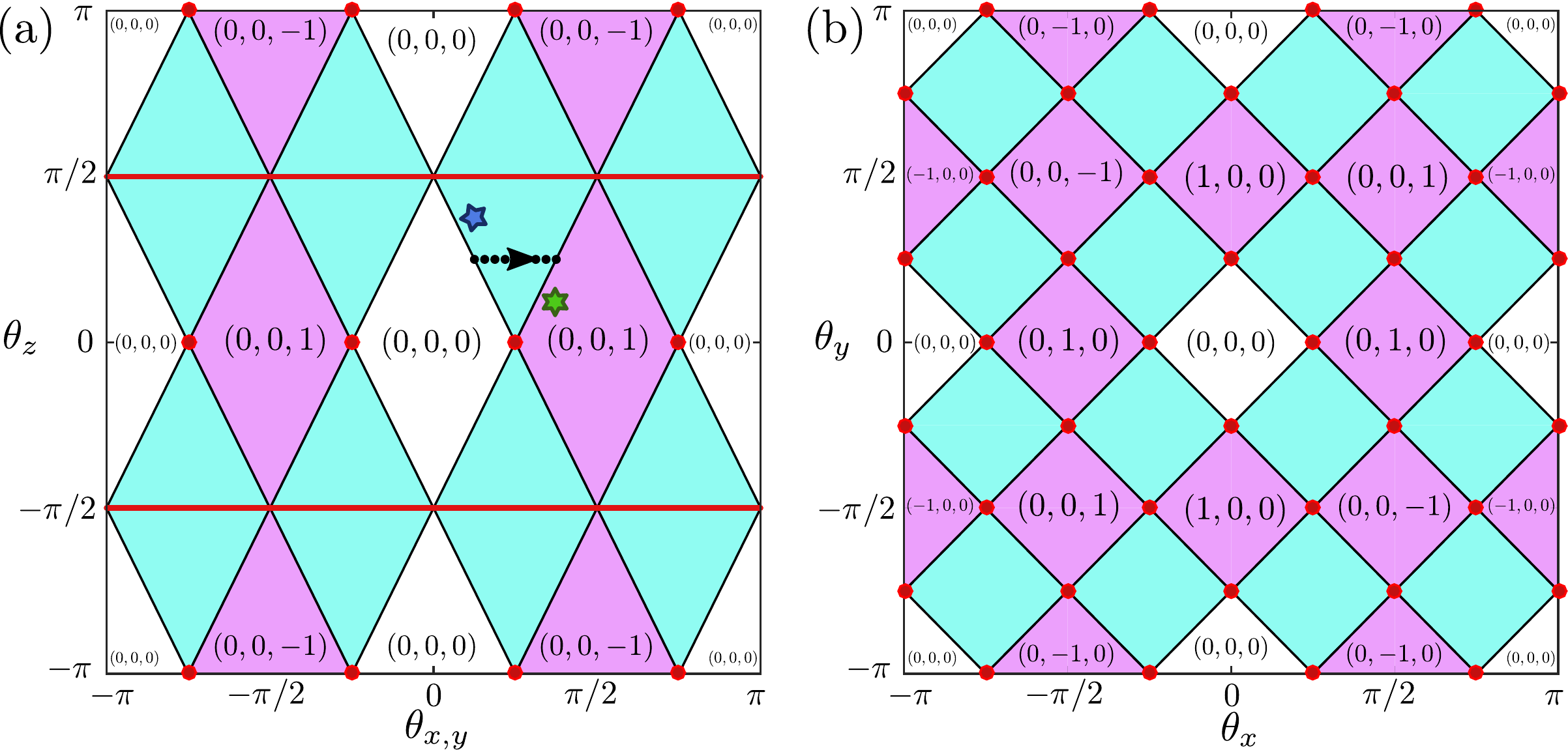}
  \caption{(Color online) Phase diagram of the 3D network for two
    different sections of the parameter space
    $\{\theta_x,\theta_y,\theta_z\}$: (a) $\theta_x=\theta_y$ and (b)
    $\theta_z=\pi/4$.  The white regions are conventional insulator
    phases; the pink regions are $\mathcal{T}$-broken weak topological
    insulator phases characterized by nonzero winding number triplets
    $(n_x,n_y,n_z)$; and the cyan regions are gapless Weyl
    phases. Along the red lines and dots, the band-touching points
    form line nodes.  The dotted line is the path producing the Weyl
    point trajectories in Fig.~\ref{fig:3D_reflection_coeff1}(d).
    Green and blue stars indicate the parameters for
    Fig.~\ref{fig:3D_reflection_coeff1} and
    Fig.~\ref{fig:3D_reflection_coeff}, respectively.}
  \label{fig:3D_phase_diagram}
\end{figure}

Fig.~\ref{fig:3D_phase_diagram} shows the topological phase diagram of
the network model, with the gapped phases classified using the pumping
invariants.  For ease of visualization, we have reduced the parameter
space to 2D by taking the section $\theta_x=\theta_y$ in
Fig.~\ref{fig:3D_phase_diagram}(a), and $\theta_z=\pi/4$ in
Fig.~\ref{fig:3D_phase_diagram}(b).  The white regions are gapped
phases with invariants $(0,0,0)$, corresponding to conventional
insulators.  The pink regions are topologically non-trivial gapped
phases, which fall into three groups with either $n_x$, $n_y$, or
$n_z$ non-zero.  The cyan regions are the Weyl phases.  Note that any
transition between different gapped phases requires passing through a
Weyl phase.

Consider the $(0,0,1)$ phase.  It has a topologically protected
surface state if the slab is taken in the $x$-$z$ or $y$-$z$
direction.  This follows from the fact that $\omega_\pm = 0$
corresponds to the Dirichlet boundary conditions for a surface state
calculation (see Section \ref{sec:Fermi_arc}).  This behavior is
reminiscent of ``weak topological insulator'' phases \cite{Fu2007},
and hence we refer to the topologically nontrivial gapped phases as
such.  (Note, however, that the ``weak topological insulator''
terminology was originally introduced in the context of $\mathcal{T}$
symmetric materials, whereas $\mathcal{T}$ is broken in this model.)
The gapped phase can be adiabatically continued to a stack of 2D
sheets, similar to the Chalker-Dohmen model \cite{ChalkerDohmen}.  In
fact, each sheet is a 2D anomalous Floquet topological insulator
\cite{Kitagawa2010,MLevin2013,Yidong2014}, which has the unique
feature that the winding-number invariant is the same in \textit{all}
quasienergy gaps \cite{Yidong2014}.  This is consistent with
above-mentioned finding that the $(n_x,n_y,n_z)$ invariant is
gap-independent.

The pumping invariant $(n_x,n_y,n_z)$ was derived in terms of the
winding numbers of the reflection phase in the 2D Brillouin zone.
However, they can also be related to the winding numbers of the Weyl
point trajectories.  We can regard the gapped phases as being
``generated'' via Weyl phases by the mutual annihilation of Weyl
points.  In this view, the different types of trajectories undertaken
by the Weyl points, between creation and annihilation, give rise to
the various topologically-distinct gapped phases \cite{Murakami2008}.

To illustrate this, consider a parameter-space trajectory from a
$(0,0,0)$ conventional insulator to a $(0,0,1)$ weak topological
insulator, transiting through a Weyl phase as indicated by the dotted
arrow in Fig.~\ref{fig:3D_phase_diagram}(a).  The trajectories of the
Weyl points under this parameter evolution are plotted in
Fig.~\ref{fig:3D_reflection_coeff1}(d).  Throughout, the Weyl points
lie in the plane $k_x = k_y$.  At the boundary with the conventional
insulator, a pair of Weyl points with opposite chiralities is
generated at $k = [0,0,0]$ with quasienergy $\phi = \pi$, and another
pair is generated at $k = [\pi,\pi,\pi]$ with quasienergy $\phi = 0$.
As the system transits through the Weyl phase, the Weyl points move
across the Brillouin zone and recombine once the system reaches the
phase boundary with the $(0,0,1)$ phase.

Based on this, we can define a net winding number
\begin{equation}
  \Delta n_\mu = - \frac{1}{2\pi} \sum_{c \in \pm 1} c \int_{0}^1 d\lambda\,
  \frac{dk^\mu_{c}}{d\lambda},
  \label{Weyl_winding}
\end{equation}
where $k^\mu_{c}(\lambda)$ denotes the $k$-space position of the Weyl
point with topological charge $c$, and $\lambda \in [0,1]$ is a path
parameter whose endpoints correspond to phase boundaries with gapped
phases.  Assuming $\lambda = 0$ is a conventional insulator, $(\Delta
n_x, \Delta n_y, \Delta n_z)$ is precisely equal to the pumping
invariant $(n_x, n_y, n_z)$ for the gapped phase at $\lambda = 1$.
This follows from the earlier observation that a Weyl point with
topological charge $c$ produces a phase singularity with winding
number $\pm c$ (on the $\pm$ surface).

A similar sort of winding behavior was previously noted by Murakami
and Kuga \cite{Murakami2008}, in the context of 3D quantum spin Hall
systems.  They found that crossing a Weyl phase between two spin Hall
insulators, such as a weak and strong topological insulator, produced
a $k$-space motion of \textit{two} pairs of Weyl points.  Rather than
wrapping around the Brillouin zone, as in the present network model,
the Weyl points encircled a time-reversal invariant momentum point,
and the net winding could be linked to a $Z_2$ topological number
\cite{Murakami2008}.

\section{Discussion}

In this paper, we have described a 3D network model exhibiting Weyl,
conventional insulator, and $\mathcal{T}$-broken weak topological
insulator phases.  The 3D model is closely related to
previously-studied 2D network models
\cite{Liang,Yidong2014,Yidong2015,Tauber}; specifically, the weak
topological insulator phases can be adiabatically continued to a stack
of decoupled 2D anomalous Floquet insulators (which are ``anomalous''
because their bandstructures are topologically non-trivial despite all
bands having zero Chern number \cite{Kitagawa2010, MLevin2013,
  Gong2014}).  Our study relied in part on the ``topological pumping''
invariant, which relates bandstructure topology to reflection
coefficients \cite{Brouwer1998,Brouwer2011,Fulga,Fulga2015}.  Though
this method was originally developed with topological insulators in
mind, we have shown that it has interesting applications to Weyl
phases: the Weyl points produce phase singularities in the reflection
spectrum, associated with the Fermi arc surface states
\cite{SMYoung,Fermi_note}.  From this, we found an intriguing
correspondence between the topology of the insulator phases, and the
$k$-space topology of Weyl point trajectories as they wind around the
Brillouin zone.  In this view, the various insulator phases are
generated through the creation and annihilation of Weyl point pairs
with different $k$-space windings.

A promising route to realizing the network model is to use coupled
microwave components, similar to the experiment reported in
Ref.~\onlinecite{Yidong2015} which realized a 2D topological pump
acting on an anomalous Floquet insulator.  The network links would be
microwave lines (e.g.~coaxial cables), with line delays fixed at $\phi
= 0$ or $\phi = \pi$ (the Weyl point quasienergies).  The nodes would
be four-port couplers, with isolators to enforce directionality.
Rather than implementing a large slab, tunable phase-shifters can
simulate Bloch boundary conditions \cite{Yidong2015}, using two sets
of phase-shifters corresponding to (e.g.) $k_x$ and $k_y$.  The key
experimental signatures would be the phase singularities, and their
$k$-space trajectories, as discussed in Section \ref{sec:Topological
  Invariants}.  Although an electromagnetic Weyl phase has previously
been realized \cite{LLu2013}, that photonic crystal-based experiment
was unable to probe the topological features of the Weyl points.  Our
proposed approach would demonstrate the topological robustness of the
Weyl points through their association with phase singularities.  This
requires measuring complex wave parameters (i.e., including phase
information), which is achievable with microwave vector network
analyzers \cite{Yidong2015,Gao}.

The network model that we have chosen to study is, in a sense, the
simplest non-trivial design that generalizes the Chalker-Coddington 2D
network model to 3D, while maintaining similar network topology in the
$x$, $y$, and $z$ directions.  We have focused on studying various
interesting properties of the disorder-free network; however, it is
worth noting that the original motivation of network models was to
provide a computationally efficient method for studying the effects of
disorder \cite{ChalkerCo}.  In future work, we intend to investigate
whether such 3D network models can be used to model disordered Weyl
semimetals.  3D network models with more complicated network
topologies may also exhibit other topological phases remaining to be
explored.  It might also be interesting to realize static
(non-Floquet) Hamiltonian models that can exhibit the phenomenon of
Weyl point trajectories winding around the Brillouin zone.

We thank W.~Hu, D.~Leykam, C. Huang, P. Chen, B.~Zhang, and L.~Lu for
helpful comments.  This research was supported by the Singapore
National Research Foundation under grant No.~NRFF2012-02, and by the
Singapore MOE Academic Research Fund Tier 3 grant MOE2011-T3-1-005.


\appendix

\section{3D network bandstructure}
\label{3Dnetwork}

This appendix describes the derivation of the evolution matrix
characterizing the network model introduced in Section \ref{sec:3D
  Network Model}.  As previously discussed, in each unit cell the
amplitudes entering the nodes are $\psi_{\mu}^\pm$, and those leaving
the nodes are $\varphi_\mu^\pm$.  For Bloch modes, the scattering
relations are given in Eq.~(\ref{eqn:scattering}).  This can be
re-arranged as
\begin{equation}
  \begin{pmatrix}
    \varphi_{\mu}^{-}\\ \varphi_{\mu}^{+}\\
  \end{pmatrix}
  =
  \begin{pmatrix}
    e^{-ik_\mu}\sin\theta_\mu&i\cos\theta_\mu\\
    i\cos\theta_\mu&e^{ik_\mu}\sin\theta_\mu\\
  \end{pmatrix}
  \begin{pmatrix}
    \psi_{\mu}^{+}\\ \psi_{\mu}^{-}\\
  \end{pmatrix}\;.
  \label{eqn:Apxscattering1}
\end{equation}
On the other hand, the wave amplitudes are also related by the fact
that traversing each link incurs a phase delay of $\phi/3$.  Thus
(referring to Fig.~\ref{fig:network}):
\begin{align}
  \begin{pmatrix}
    \psi_{z}^{+}\\ \psi_{z}^{-}\\
  \end{pmatrix}
  &=e^{i\frac{\phi}{3}}
  \begin{pmatrix}
    \varphi_{y}^{-}\\ \varphi_{y}^{+}\\
  \end{pmatrix} \\
  \begin{pmatrix}
    \psi_{y}^{+}\\ \psi_{y}^{-}\\
  \end{pmatrix}
  &=e^{i\frac{\phi}{3}}
  \begin{pmatrix}
    \varphi_{x}^{-}\\ \varphi_{x}^{+}\\
  \end{pmatrix}\\\begin{pmatrix}
  \psi_{x}^{+}\\ \psi_{x}^{-}\\
  \end{pmatrix}
  &=e^{i\frac{\phi}{3}}
  \begin{pmatrix}
    \varphi_{z}^{-}\\ \varphi_{z}^{+}\\
  \end{pmatrix}\;.
\label{eqn:Apxscattering2}
\end{align}
Combining Eqs.~(\ref{eqn:Apxscattering1})--(\ref{eqn:Apxscattering2})
gives $U\Psi = e^{-i\phi} \Psi$, where
\begin{align}
  \begin{aligned}
    U &= U_z U_y U_x, \\
    U_\mu &\equiv \begin{pmatrix}
      e^{-ik_\mu}\sin\theta_\mu&i\cos\theta_\mu\\
      i\cos\theta_\mu&e^{ik_\mu}\sin\theta_\mu\\
    \end{pmatrix}.
  \end{aligned}
  \label{eqn:Umatrix}
\end{align}
From its eigenvalues, we find the quasienergies
\begin{align}
  \begin{aligned}
  \phi=\pm\cos^{-1}\Big[&\cos(k_x+k_y+k_z)\sin\theta_x\sin\theta_y\sin\theta_z
    \\
&    -\cos k_x\sin\theta_x\cos\theta_y\cos\theta_z
    \\
&    -\cos k_y\cos\theta_x\sin\theta_y\cos\theta_z
    \\
&    -\cos k_z\cos\theta_x\cos\theta_y\sin\theta_z\Big].
  \end{aligned}
\end{align}
The band-crossing points are determined by the degeneracies of the
multiple-valued arc cosine, so either $\phi = \pi$ or $\phi = 0$.  For
$\phi = \pi$, degeneracies occur at $\pm\bar{k}_\mu$, where
\begin{align}
  \begin{aligned}
  \bar k_x &= s_x \cos^{-1}\left(\frac{- \cos^2\theta_x + \cos^2\theta_y + \cos^2\theta_z}{2\sin\theta_x\cos\theta_y\cos\theta_z}\right) \\
    \bar k_y &= s_y \cos^{-1}\left(\frac{\cos^2\theta_x - \cos^2\theta_y + \cos^2\theta_z}{2\cos\theta_x\sin\theta_y\cos\theta_z}\right) \\
    \bar k_z &= - s_z\cos^{-1}\left(\frac{\cos^2\theta_x+\cos^2\theta_y-\cos^2\theta_z}{2\cos\theta_x\cos\theta_y\sin\theta_z}\right),
  \end{aligned}
  \label{eqn:kbar}
\end{align}
where $s_\mu \equiv \text{sgn}(\sin2\theta_\mu)$.  For $\phi=0$,
degeneracies occur at $\pm(\bar{k} - [\pi,\pi,\pi])$.

Using Eqs.~(\ref{eqn:Umatrix})--(\ref{eqn:kbar}), we expand the
evolution matrix in the vicinity of the Weyl point $\bar{k}$, as $U
\approx e^{-iH_{\text{eff}}}$.  To leading order in the displacement
$\kappa = k - \bar{k}$, we find that $H_{\text{eff}} =
\nu_{ij}\kappa_{i}\sigma_j$, with coefficients
\begin{align*}
&\begin{array}{ll}
\nu_{xx} &= -     g_{yz} \sin\bar{k}_x\\
\nu_{xy} &= \;\;\;g_{yz} \cos\bar{k}_x\\
\nu_{xz} &= \;\;\;g_{yz} \tan\theta_x\\
\end{array},\quad
\begin{array}{ll}
\nu_{yx} &= -     g_{xz} \sin\bar{k}_y\\
\nu_{yy} &= -     g_{xz} \cos\bar{k}_y\\
\nu_{yz} &= \;\;\;g_{xz} \tan\theta_y\\
\end{array} \\
&\;\nu_{zx}\; = -     \left[f_x \sin(\bar{k}_y+\bar{k}_z) + f_y \sin(\bar{k}_x+\bar{k}_z) \right]\\
&\;\nu_{zy}\; = -     \left[f_x \cos(\bar{k}_y+\bar{k}_z) - f_y \cos(\bar{k}_x+\bar{k}_z) \right]\\
&\;\nu_{zz}\; = - \cos\theta_x\cos\theta_y\sin\theta_z \cos \bar{k}_z \\
& \quad\qquad - \sin\theta_x\sin\theta_y\sin\theta_z \cos(\bar{k}_x+\bar{k}_y+\bar{k}_z),
\end{align*}
where
\begin{align}
  \begin{aligned}
  g_{\mu\nu} &\equiv  \sin\theta_x\sin\theta_y\sin\theta_z\left(\frac{\cos\bar{k}_\nu}{\tan\theta_\mu}
    + \frac{\cos\bar{k}_\mu}{\tan\theta_\nu}\right)\\
  f_{\mu} &\equiv \frac{\sin\theta_x\sin\theta_y\sin\theta_z}{\tan\theta_\nu}.
  \end{aligned}
  \label{fg}
\end{align}
From this, Eq.~(\ref{eqn:detnu}) can be derived via direct
substitution.

We have thus far assumed that the coupling matrices at the nodes have
the simple form given in Eq.~(\ref{eqn:scattering}), corresponding to
couplers with $180^\circ$ rotational symmetry \cite{Liang,Liang2014}.
The discussion can be generalized to arbitrary $2\times2$ unitary
coupling matrices of the form
\begin{equation}
  \mathcal{U}_c
    =e^{i\Phi_3^\mu}
  \begin{pmatrix}
    \sin\theta_\mu e^{-i(\Phi_1^\mu + \Phi_2^\mu)}
    &i\cos\theta_\mu e^{i(\Phi_1^\mu - \Phi_2^\mu)} \\
    i\cos\theta_\mu e^{-i(\Phi_1^\mu - \Phi_2^\mu)}
    &\sin\theta_\mu e^{i(\Phi_1^\mu + \Phi_2^\mu)}\\
  \end{pmatrix},
\end{equation}
where $\{\Phi_1^{\mu},\Phi_2^{\mu},\Phi_3^{\mu}\}$ are additional
Euler angles which had previously been ignored (set to zero).  With
this generalization, the preceding results
(\ref{eqn:Apxscattering1})--(\ref{fg}) still hold, subject to the
replacement
\begin{align}
  \begin{aligned}
    \phi &\rightarrow \phi + \Phi_3^x+\Phi_3^y+\Phi_3^z \\ 
    k_x &\rightarrow k_x - \Phi_1^x-\Phi_2^x-\Phi_1^y+\Phi_2^y+\Phi_1^z-\Phi_2^z \\
    k_y &\rightarrow k_y + \Phi_1^x-\Phi_2^x-\Phi_1^y-\Phi_2^y-\Phi_1^z+\Phi_2^z\\
    k_z &\rightarrow k_z - \Phi_1^x+\Phi_2^x+\Phi_1^y-\Phi_2^y-\Phi_1^z-\Phi_2^z.
  \end{aligned}
\end{align}
In other words, the Euler angles $\Phi_1^\mu$ and $\Phi_2^\mu$
translate the quasienergy bandstructure in $k$-space, whereas the
Euler angle $\Phi_3^\mu$ translates it in $\phi$ (thus, for
$\Phi_2^\mu \ne 0$, Weyl points would no longer occur at $\phi = 0$
and $\phi = \pi$).  Such translations do not, however, alter the
topological properties of the network bandstructure.

\section{Bandstructures of finite-thickness slabs}
\label{Tmatrix}

To obtain the bandstructure of a finite-thickness slab (Section
\ref{sec:Fermi_arc}), and the topological pumping invariants (Section
\ref{sec:Topological Invariants}), we need to calculate the transfer
matrix for crossing the network in each direction.  Consider the
transfer matrix in $z$ (the other two transfer matrices are worked out
similarly): the network is infinite and periodic in $x$ and $y$
directions, with quasimomenta $(k_x,k_y)$.  The transfer matrix across
one unit cell in the $z$ direction is defined by
\begin{equation}
  M
  \begin{pmatrix}
    \varphi_{z}^{-}(j)\\ \psi_{z}^{-}(j)\\
  \end{pmatrix} =
  \begin{pmatrix}
    \varphi_{z}^{-}(j+1)\\ \psi_{z}^{-}(j+1)\\
  \end{pmatrix},
  \label{eqn:transfer2a}
\end{equation}
where $j$ and $j+1$ are adjacent cell indices along $z$.  We can find
$M$ using the network model definitions, as follows.  Firstly, based
on Eq.~(\ref{eqn:scattering}), we can relate the amplitudes at the
bottom of cell $j+1$ to those at the top of cell $j$ by
\begin{align}
\begin{aligned}
  M'
  \begin{pmatrix}
    \psi_{z}^{+}(j) \\ \varphi_{z}^{+}(j)
  \end{pmatrix} &=
  \begin{pmatrix}
    \varphi_{z}^{-}(j+1)\\ \psi_{z}^{-}(j+1)
  \end{pmatrix}, \\ \mathrm{where}\;\;
  M' &\equiv \begin{pmatrix}
    \csc\theta_z & i\cot\theta_z \\
    -i \cot\theta_z & \csc\theta_z
  \end{pmatrix}.
\end{aligned}
\label{eqn:Apxtransfer1}
\end{align}
Next, we use
Eqs.~(\ref{eqn:Apxscattering1})--(\ref{eqn:Apxscattering2}) to relate
$[\varphi_{z}^{+}(j), \psi_{z}^{+}(j)]^{\text{T}}$ to
$[\varphi_{z}^{-}(j), \psi_{z}^{-}(j)]^{\text{T}}$, the amplitudes at
the bottom of cell $j$.  This introduces the phase delay $\phi/3$.
The result is
\begin{equation*}
  M = \frac{1}{\chi}
  \begin{pmatrix}
    e^{i\phi}+ \xi \cos\theta_z &
    i \left(e^{-i\phi}\cos\theta_z+ \xi^*\right) \\
    -i \left(e^{i\phi}\cos\theta_z+ \xi\right) &
    e^{-i\phi}+ \xi^*\cos\theta_z
  \end{pmatrix},
\end{equation*}
where
\begin{align*}
  \xi &\equiv e^{-ik_x}\sin\theta_x\cos\theta_y+e^{ik_y}\cos\theta_x\sin\theta_y \\
  \chi &\equiv \left[e^{i(k_x+k_y)}\sin\theta_x\sin\theta_y-\cos\theta_x\cos\theta_y\right]\sin\theta_z.
\end{align*}
From the one-cell transfer matrix $M$, we can calculate the transfer
matrix across a stack of $N$ cells:
\begin{equation}
  \mathcal{M}(N)=\left(M'\right)^{-1} M^N.
\end{equation}
This transfer matrix depends on the coupling parameters
$\{\theta_\mu\}$, the transverse quasimomenta $(k_x,k_y)$, and the
quasienergy $\phi$.

\begin{figure}
\centering
\includegraphics[width=\columnwidth]{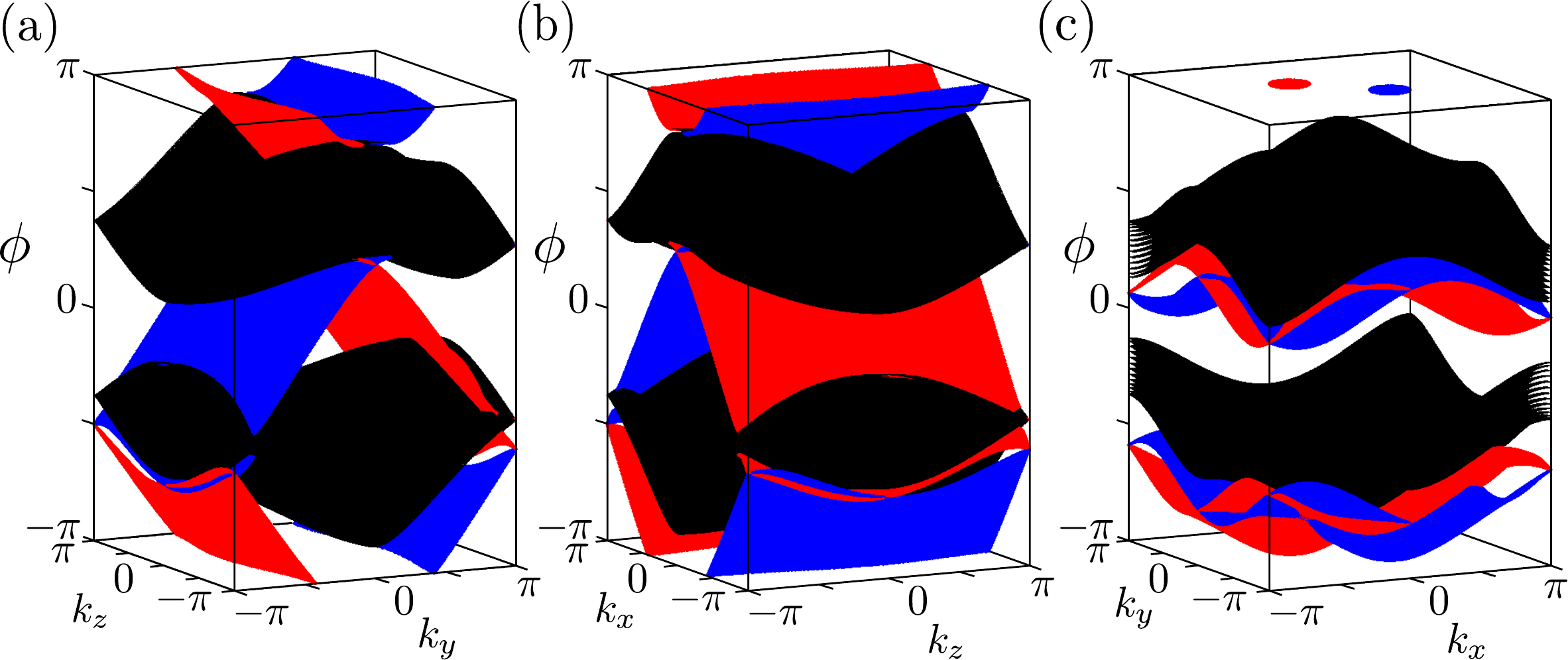}
\caption{(Color online) Quasienergy bandstructure for the 3D network
  in various slab geometries: (a) $y$-$z$, (b) $x$-$z$, and (c)
  $x$-$y$.  In all cases, the slab thickness is $N=16$, and the
  couplings are $\theta_x=\theta_y=3\pi/8$ and $\theta_z=\pi/8$,
  indicated by the green star in the phase diagram, i.e.~a $(0,0,1)$
  gapped phase. }
\label{fig:weakti_surface_spectrum}
\end{figure}

To calculate the bandstructure of the slab, we search numerically for
combinations of $\{\phi, k_x, k_y\}$ such that $[1,1]^{\text{T}}$ is
an eigenvector of $\mathcal{M}$, which corresponds to the Dirichlet
boundary conditions described in Section \ref{sec:Fermi_arc}.  The
results are as shown in Fig.~\ref{fig:3D_surface_spectrum} for the
Weyl phase, revealing the existence of Fermi arc surface states.  For
contrast, the bandstructure in a weak topological insulator phase is
shown in Fig.~\ref{fig:weakti_surface_spectrum}.

We can also use $\mathcal{M}(N)$ to derive the scattering matrix
\begin{equation}
  \mathcal{S} = \frac{1}{\mathcal{M}_{11}} \begin{pmatrix}
   \mathcal{M}_{21} & 1 \\  \det(\mathcal{M}) & -\mathcal{M}_{12}
  \end{pmatrix},
\end{equation}
which relates the wave amplitudes that are entering the lower and
upper surfaces of the slab to the outgoing wave amplitudes.  From
this, we can calculate the winding-number variant described in Section
\ref{sec:Topological Invariants}.

\section{Discrete-time quantum walks}
\label{sec:quantumwalk}

The 3D network model is described by a unitary matrix, $U(k, \theta_x,
\theta_y, \theta_z)$, which is determined by the node couplings and
the connectivity of the network links. The same matrix can be derived
from a discrete-time quantum walk (DTQW), i.e.~the discrete-time
quantum dynamics generated from a stepwise time-dependent Hamiltonian.
DTQWs can be implemented using ultra-cold atoms in optical lattices
\cite{QWrealization}. In 2D, DTQWs can also be realized using coupled
optical waveguide arrays \cite{waveguides}, but this requires using a
third spatial dimension (the waveguide axis) to play the role of time,
so 3D quantum walks can not be implemented this way.

\begin{figure}
  \centering
  \includegraphics[width=\columnwidth]{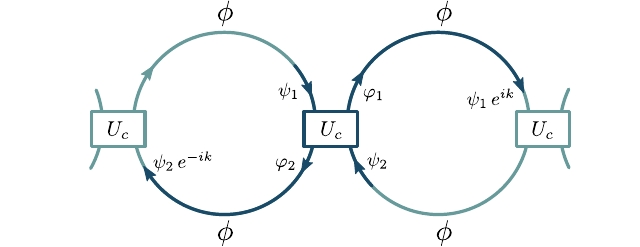}
  \caption{(Color online) Schematic of a 1D network model.}
  \label{fig:onednetwork}
\end{figure}

To begin, we consider the 1D network model shown in
Fig.~\ref{fig:onednetwork}.  At a typical node in the network, there
is an input-output relation
\begin{equation}
  \mathcal{U}_c \begin{pmatrix}\psi_1 \\ \psi_2 \end{pmatrix} = \begin{pmatrix}\varphi_1 \\ \varphi_2 \end{pmatrix},
\end{equation}
where $\psi_1$, $\psi_2$, $\varphi_1$, and $\varphi_2$ are the complex
wave amplitudes at the positions indicated in the figure, and
$\mathcal{U}_c$ is a fixed $2\times2$ unitary coupling matrix.  For a
Bloch state with quasimomentum $k$, the wave amplitudes are further
related by $\varphi_1 = e^{-i\phi} e^{ik} \psi_1$ and $\varphi_2 =
e^{-i\phi} \, e^{-ik} \psi_2$, where $\phi$ is the link delay.  The
$k$-space evolution matrix is
\begin{equation}
  U(k) = e^{-ik\sigma_z} \, \mathcal{U}_c.
  \label{oneduk}
\end{equation}
Note that $\mathcal{U}_c$ is independent of $k$.  We can map $U(k)$
onto the evolution matrix for a DTQW for a particle in a 1D lattice,
with internal ``spin up'' ($\uparrow$) and ``spin down''
($\downarrow$) degrees of freedom.  The DTQW is divided into two
steps.  In the first step, we apply $U_c$ on every lattice site.  In
the second step, we perform a spin-dependent translation that moves
$\uparrow$ one site to the left, and $\downarrow$ one site to the
right:
\begin{multline}
  \sum_j |j-1\rangle\langle j|\otimes|\uparrow\;\rangle\langle\;\uparrow|+|j+1\rangle\langle j|\otimes|\downarrow\;\rangle\langle\;\downarrow| \\
  = \int_{-\pi}^{\pi}dk\; e^{-ik\sigma_z}\otimes|k\rangle\langle k|\;.
  \label{stepping}
\end{multline}
Then, for each $k$, the evolution operator over one period of the DTQW
reduces to Eq.~(\ref{oneduk}).

We can similarly map our 3D network model to a 3D DTQW.  Start from
the scattering at a given node, which is described by the unitary
scattering relation (\ref{eqn:Apxscattering1}).  To relate this to the
quantum walk, observe that
\begin{align}
  \begin{aligned}
  \begin{pmatrix}e^{-ik_{\mu}} & 0\\0 & 1\end{pmatrix}
    &=e^{-ik_{\mu}/2}e^{-ik_{\mu}\sigma_{z}/2}\\
    \begin{pmatrix}1 & 0\\0 & e^{ik_{\mu}}\end{pmatrix}
    &=e^{ik_{\mu}/2}e^{-ik_{\mu}\sigma_{z}/2}\\
    \begin{pmatrix}\sin\theta_{\mu} & i\cos\theta_{\mu}\\i\cos\theta_{\mu} & \sin\theta_{\mu}\end{pmatrix}
    &=e^{-i(\theta_{\mu}-\frac{\pi}{2})\sigma_{x}}.
  \end{aligned}
\end{align}
We can combine these operations to obtain
\begin{equation}
  U_{\mu}
  = e^{-ik_{\mu}\sigma_{z}/2}e^{-i(\theta_{\mu}-\frac{\pi}{2})\sigma_{x}}e^{-ik_{\mu}\sigma_{z}/2},
\end{equation}
where $U_\mu$ has the form given in (\ref{eqn:Umatrix}).  Hence, we
can generate the Floquet evolution matrix $U = U_zU_yU_x$.  The DQTW
protocol thus consists of three steps, one for each direction
$\{x,y,z\}$, where each step consists of two spin rotations and one
spin-dependent translation along the chosen direction.

Other periodic network models with more complicated configurations can
also be mapped onto DTQWs, but the mapping may require more than two
internal degrees of freedom.  Consider a periodic network of arbitrary
dimension, where each unit cell contains an arbitrary configuration of
coupling nodes, joined by links of equal phase delay $\phi$, with some
of the links connecting to adjacent unit cells.  As described in
Ref.~\onlinecite{Yidong2014}, the whole set of node couplings in one
unit cell can be described by
\begin{equation}
  \mathcal{U}_c\, \vec{\psi} = \vec{\varphi},
\end{equation}
where $\vec{\psi} = [\psi_1, \dots, \psi_N]$ and $\vec{\varphi} =
[\varphi_1, \dots, \varphi_N]$ are vectors of wave amplitudes that are
incoming and outgoing from the nodes, and $\mathcal{U}_c$ is an
$N\times N$ unitary matrix.  The $\vec{\psi}$ and $\vec{\varphi}$
vectors can always be arranged so that, for each $n$, $\varphi_n$ and
$\psi_n$ lie on the opposite ends of equivalent links, either in the
same unit cell or another unit cell (refer again to
Fig.~\ref{fig:onednetwork} for a 1D example).  Then we can write
\begin{equation}
  \vec{\varphi} = e^{-i\phi} \mathcal{D}(\vec{k}) \, \vec{\psi},
\end{equation}
where $\mathcal{D}(\vec{k})$ is a diagonal matrix where each diagonal
element has the form $\exp(i\vec{k}\cdot \vec{d}_n)$, and $\vec{d}_n$
is a lattice displacement vector representing the lattice displacement
for link $n$.  Hence, the periodic network can be described by a
$k$-space unitary evolution matrix $U(k) = [\mathcal{D}(k)]^{-1}\,
\mathcal{U}_c$.  This can be implemented as a DTQW, with
$[\mathcal{D}(k)]^{-1}$ realized using a translation operation
analogous to Eq.~(\ref{stepping}).


\begin{thebibliography}{99}
\bibitem{Fermi_note} Although we refer to ``Fermi arcs", in
  electromagnetic bandstructures the particles do not obey Fermi
  statistics.
\end{thebibliography}
\end{document}